# Storage-ring Electron Cooler for Relativistic Ion Beams[*]

F. Lin[†], Y.S. Derbenev, D. Douglas, J. Guo, G. Krafft, V.S. Morozov, Y. Zhang, Jefferson Lab, Newport News, VA USA
R.P. Johnson, MuPlus, Inc., Newport News, VA USA

May 9, 2016

## Abstract

Application of electron cooling at ion energies above a few GeV has been limited due to reduction of electron cooling efficiency with energy and difficulty in producing and accelerating a high-current high-quality electron beam. A high-current storage-ring electron cooler offers a solution to both of these problems by maintaining high cooling beam quality through naturally-occurring synchrotron radiation damping of the electron beam. However, the range of ion energies where storage-ring electron cooling can be used has been limited by low electron beam damping rates at low ion energies and high equilibrium electron energy spread at high ion energies. This paper reports a development of a storage ring based cooler consisting of two sections with significantly different energies: the cooling and damping sections. The electron energy and other parameters in the cooling section are adjusted for optimum cooling of a stored ion beam. The beam parameters in the damping section are adjusted for optimum damping of the electron beam. The necessary energy difference is provided by an energy recovering SRF structure. A prototype linear optics of such storage-ring cooler and initial tracking simulations are presented and some potential issues such as coherent synchrotron radiation and beam break up are discussed.

## Introduction

Cooling of ion beams to achieve a significant reduction of six dimensional beam emittance is crucial to deliver high luminosities over a broad center of mass energy range in a collider [1, 2]. Cooling allows small transverse beam sizes at the interaction point and small $\beta^*$ with crab crossings and short bunch lengths. Cooling also counteracts intra-beam scattering (IBS) emittance degradation and, as a consequence, extends luminosity lifetime. Unlike electron beams that have intrinsic synchrotron radiation damping, ion beams have no radiation damping in low or medium energy ranges (a few to a few hundred GeV). An efficient ion damping mechanism must be introduced.

---

An electron-beam-based cooling system is one of the most promising approaches to cool ion beams [3, 4]. Considering that most colliders are designed to have large bunch charge and high collision frequency to boost the luminosity, such a cooling facility should be capable of providing high energy (up to a few tens of MeV) and high current (up to a few Amperes) electron beams with a high bunch repetition rate and good quality (small emittance and energy spread). To obtain such high energy and high current electron beams, an energy recovering linac (ERL) is required to boost the electron beam and recover the large beam power. To reduce the challenge on the lifetime of the source, a circulator ring, where the electron beam circulates up to a few tens of turns while continuously cooling the ion beam, was proposed [1]. It reduces the average current of the electron beam from the cathode and ERL by a factor of the number of circulations. However, this facility demands several key technology developments, such as magnetized bunched beam electron gun [5] and injector, and a fast kicker [1] for switching bunches into and out of the circulator ring.

These constraints will disappear or diminish significantly if a storage ring electron cooling methodology is applied [6, 7]. High electron beam current can be obtained by stacking electron current and having the electron bunches circulate in a storage ring. High cooling beam quality could be obtained through naturally-occurring synchrotron radiation damping counteracting the heating of the electron beam by the intra- and inter-beam scattering. If needed, damping wigglers can be used to enhance the damping effect. However, the range of ion energies where storage-ring electron cooling can be used has been limited by low electron beam damping rates at low energies, which is proportional to the 3$^{rd}$ power of the beam energy $E$ (at fixed bend radii as given in Eq. (1)), and large equilibrium electron energy relative spread at high energies due to quantum radiation, which is proportional to the 1$^{st}$ power of the beam energy as given in Eq. (2).

The electron beam damping rate is given by

$$\tau_i^{-1} = \frac{J_i}{2ET_o} \cdot \frac{C_r E^4}{2\pi} \oint \frac{ds}{\rho^2}, \tag{1}$$

where $i=x$ (horizontal), $z$ (vertical), $l$ (longitudinal), $J_i$ is the radiation damping partition coefficients, $C_r = 8.85 \times 10^{-5}$ m/(GeV)$^3$, $T_o$ is the revolution period and $\rho$ is the local bending radius.

The electron equilibrium energy spread is given by

$$\frac{\sigma_E}{E} = \sqrt{\frac{C_q \gamma^2}{J_l \langle 1/\rho^2 \rangle} \langle 1/\rho^3 \rangle}, \tag{2}$$

where $C_q = 3.83 \times 10^{-13}$ m, $J_l$ is the longitudinal radiation damping partition coefficient and $\rho$ is the local bending radius.

# Dual-energy Storage Ring Cooler

## *Concept*

We propose a novel concept that greatly expands the range of applicability of storage-ring electron cooling by boosting the electron beam energy in the first damping section of the cooler, where damping wigglers decrease the damping time and control the energy spread, then deaccelerating the electron beam energy to match the velocity of the ion beam in the second cooling section. Figure 1 shows a schematic drawing of such a dual-energy storage ring cooler. This scheme is also applicable at high ion energies where the damping rate is fast but the electron energy spread is too high. The electron beam energy then has to be lowered in the damping section to keep the momentum spread under control and reaccelerated back in the cooling section.

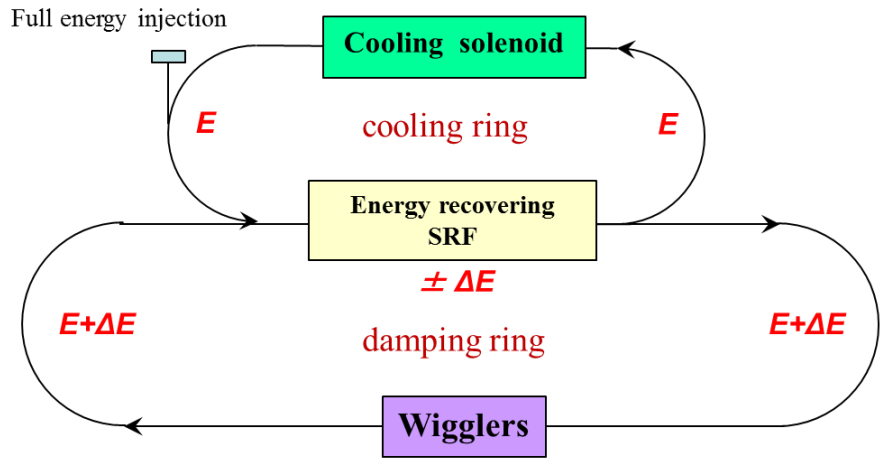

Figure 1: Schematic drawing of a dual-energy storage ring cooler.

Such concept of storage-ring electron cooling is being developed, in which a cooler consists of two sections with significantly different energies: the cooling and damping sections. These two sections are connected by an energy recovering SRF structure that provides the necessary energy difference. The electron energy, bunch length, and energy spread in the cooling section are determined for optimum cooling of a stored ion beam. In addition, matched optics in the cooling solenoid are designed to improve cooling efficiency [8]. The beam parameters in the damping section where the electron beam is boosted to a higher energy are adjusted for optimum damping of the electron beam. Wigglers are introduced to enhance the damping effect and the optics are designed to reach the required equilibrium emittance and energy spread. Since round canonical-angular-momentum (CAM) dominated electron beams in the cooling section, where the beam energy is relatively low, can suppress space charge induced instabilities, round-to-flat and flat-to-round beam transforms are performed at the entrance and exit of the damping section.

## *Technical Approach*

Table 1 shows high level design parameters for a cooling electron beam in the Jefferson Lab Electron Ion Collider (JLEIC) under development at Jefferson Lab [9]. Electron beam energies from 20 to 55 MeV are required in order to cool the ion beam from 20 up to 100 GeV for different physics study scenarios. To optimize the cooling performance, an rms bunch length of 3 cm and an energy spread of $1\times10^{-4}$ are chosen for the cooling electron bunch. Large normalized emittance is required to suppress heating effects, such as space charge, IBS, etc., that induce beam instability at low energy.

Table 1: Design parameters of the cooling electron beam in the JLEIC at Jefferson Lab.

| Parameter | Unit | |
|---|---|---|
| Energy | MeV | 20-55 |
| Current | A | 0.5-1 |
| Repetition rate | MHz | 476 |
| rms bunch length | cm | 3 |
| Normalized emittance* | μm-rad | 100-500 |
| Energy spread | $10^{-4}$ | 1 |
| Solenoid field | T | 2 |
| Solenoid length | m | 20-30 |

*This is the large canonical emittance (drift emittance) determining the beam size in the cooling section.

As an example, a complete scheme with a consistent set of parameters at 55 MeV cooling electron beam energy is developed, as shown in Figure 2. The electron beam information, in terms of bunch length and energy spread, along the storage ring is indicated by Roman numerals. The storage ring cooler has two sections or rings, a 55 MeV cooling ring and a 155 MeV damping ring. The beam is CAM-dominated round in the cooling ring in order to suppress space charge effects at low energy. The bunch length and energy spread of 3 cm and $1\times10^{-4}$, respectively, before (xi) and after (i) the cooling solenoid are the design parameters. The round beam in the cooling section is CAM-dominated, which means that one of the canonical transverse emittances is much larger than the other one. The larger emittance determines the beam size in the cooler solenoid and therefore must be large enough so that the cooling electron beam completely encompasses the cooled ion beam. Moreover, this large emittance suppresses space charge and, to some extent, IBS effects. The smaller emittance is made as small as possible, limited by non-linearities, collective effects, and synchrotron radiation.

Energy chirping after (i) is introduced to create a correlation between particle longitudinal positions and energies resulting in a bunch length of 3 cm and energy spread of $3\times10^{-4}$ at (ii). Such an electron bunch will be subsequently compressed by a particular arc optics design to reach a small bunch length of 1 cm at (iii). Such a short bunch length is preferred in the SRF structure in order to attain a very small momentum spread during acceleration.

In the SRF structure, the electron beam is boosted to a higher energy of 155 MeV and transported to the damping ring. A round-to-flat beam transform is performed so that the electron beam is flat in the high energy damping ring. Because of the energy gain by almost a factor of 3 in the SRF, adiabatic damping leads to a three-fold reduction of the energy spread, from $3\times10^{-4}$ at (iii) to $1\times10^{-4}$ at (iv).

Energy chirping is introduced again at (v), and the beam is further compressed using the arc to a short bunch length of 0.3 cm at (vi) to minimize the longitudinal emittance in the wiggler section while maximizing the energy spread. This is needed because the equilibrium energy spread due to the wiggler is independent of the bunch length (see Eq. (2)). The longitudinal emittance in the wiggler then determines the equilibrium longitudinal emittance in the rest of the ring. Therefore, it is beneficial to minimize the bunch length in the wiggler.

The electron beam passes through the wigglers to enhance the damping effect. The linear optics in the wiggler section is designed to attain the required equilibrium emittance and energy spread.

After decompression in another arc of the damping ring, the bunch length and energy spread of the electron beam return back to 1 cm and $3\times10^{-4}$ at (vii), the same as at (v). Applying energy dechirping at (viii) and performing a flat-to-round beam transform, the round electron beam with a short bunch length of 1 cm and small energy spread of $1\times10^{-4}$ will be deaccelerated in the energy recovering SRF structure to 55 MeV. Since the energy is lower by about a factor of 3 than that in the damping ring, the energy spread is increased by about a factor of 3 at (ix). After decompression in the cooling ring arc and energy dechirping at (x), the beam at (xi) returns back to the required characteristics with a bunch length of 3 cm and energy spread of $1\times10^{-4}$ before entering the cooling solenoid at (xi).

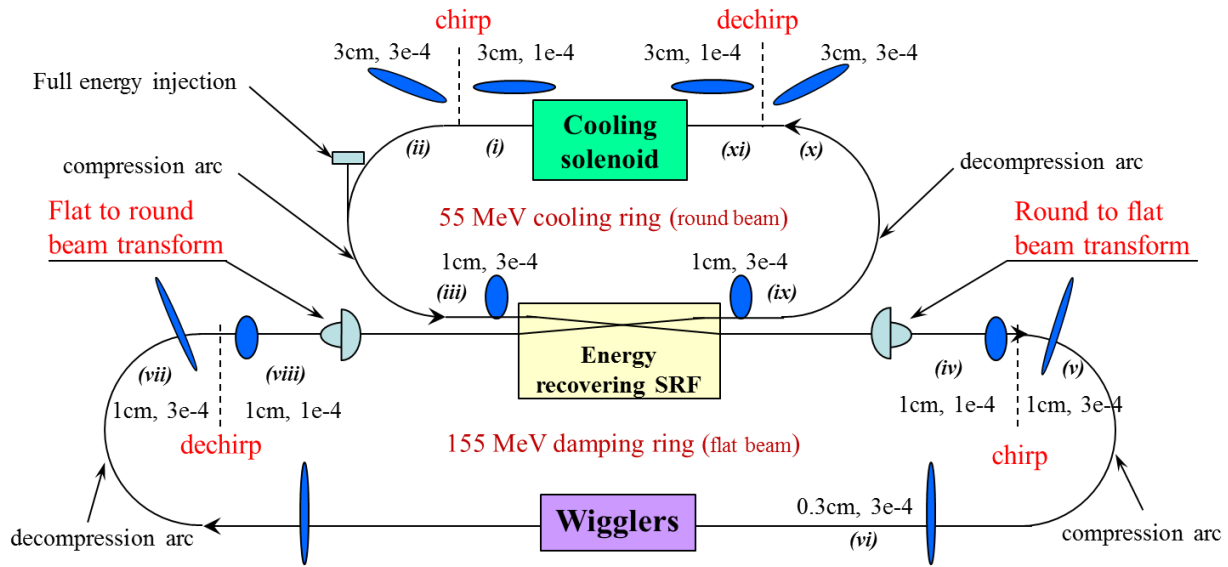

Figure 2: A complete scheme on the beam transport in the whole storage-ring electron cooler. The electron beam information, in terms of bunch length and energy spread, along the storage ring is indicated by Roman numerals.

*Linear Optics Design*

The optics of a prototype storage-ring electron cooler for the JLEIC following the aforementioned design concept is illustrated in Fig. 3. The optics of the damping ring arcs, as shown in Fig. 4, are designed with a zero-dispersion in one of the two ends to construct a dispersion-free straight for machine elements, such as SRF structures, beam transforms, etc., and non-zero dispersion leakage at the other end to introduce the dispersion in the damping wiggler section for controlling the equilibrium emittance and energy spread. Figure 5 shows the lattice functions of the whole damping wiggler section (left) and one damping wiggler unit (right). The wigglers are modeled as seven 8 m sections each consisting of a sequence of 20 cm long dipoles. The cooling ring arcs shown in Fig. 6 are designed as double bend achromat (DBA) type lattices with zero dispersion at both ends to obtain a dispersion-free straights for both cooling and RF sections.

The prototype design shown above is just an example of the storage cooler ring setup pointing out some considerations that must go into the cooler design. Clearly, finding an optimal set of parameters is not trivial. They are determined by interplay of collective, non-linear, synchrotron and imperfection effects. Future work involves finding a consistent set of parameters that provide optimal cooler performance.

The equilibrium electron beam emittances will be determined primarily by a balance of the IBS rate in the cooling "ring" and the synchrotron radiation damping rates in the damping "ring". Using a CAM-dominated (equivalent to "magnetized") round beam in the cooling section should lower the IBS and space charge effects while preserving the cooling efficiency. The initial estimates indicate that the beam parameters in the storage ring cooler can meet or exceed the requirements specified in Table 1. The beam parameters may need more quantitative studies.

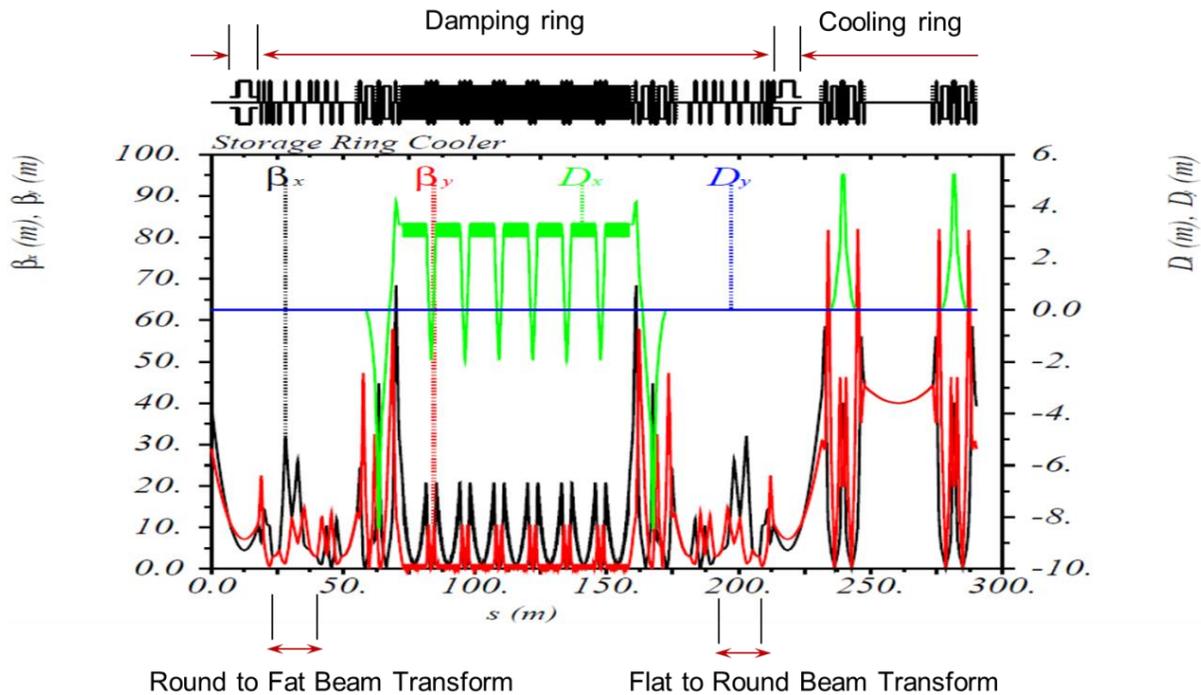

Figure 3: Optics of a prototype storage-ring electron cooler.

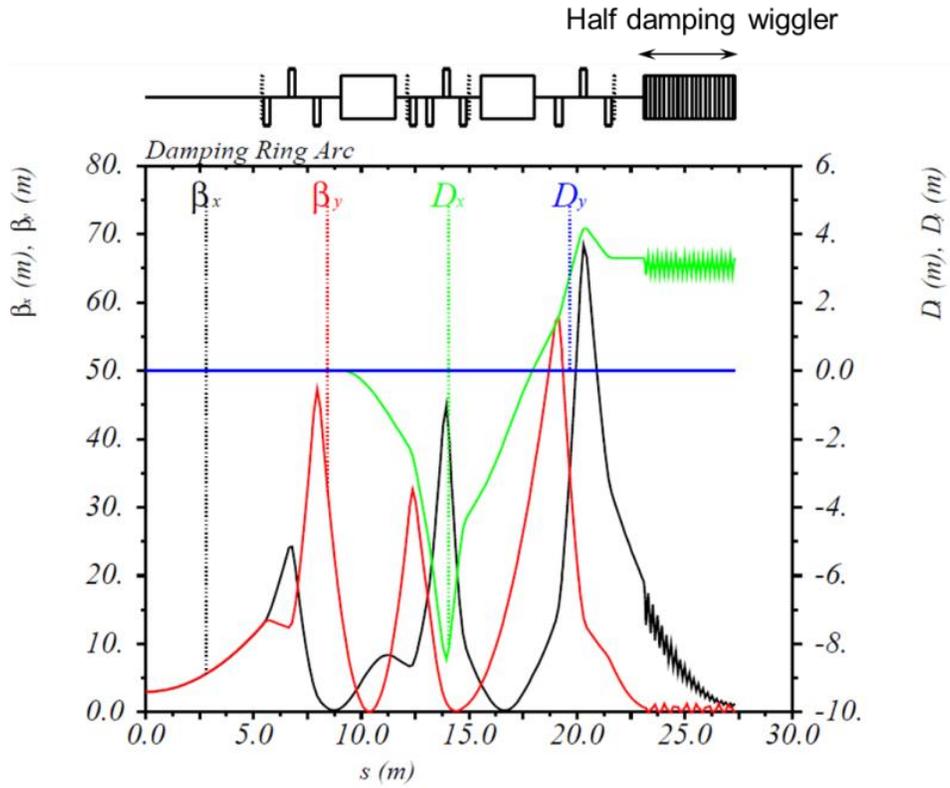

Figure 4: Optics functions of the damping ring arcs.

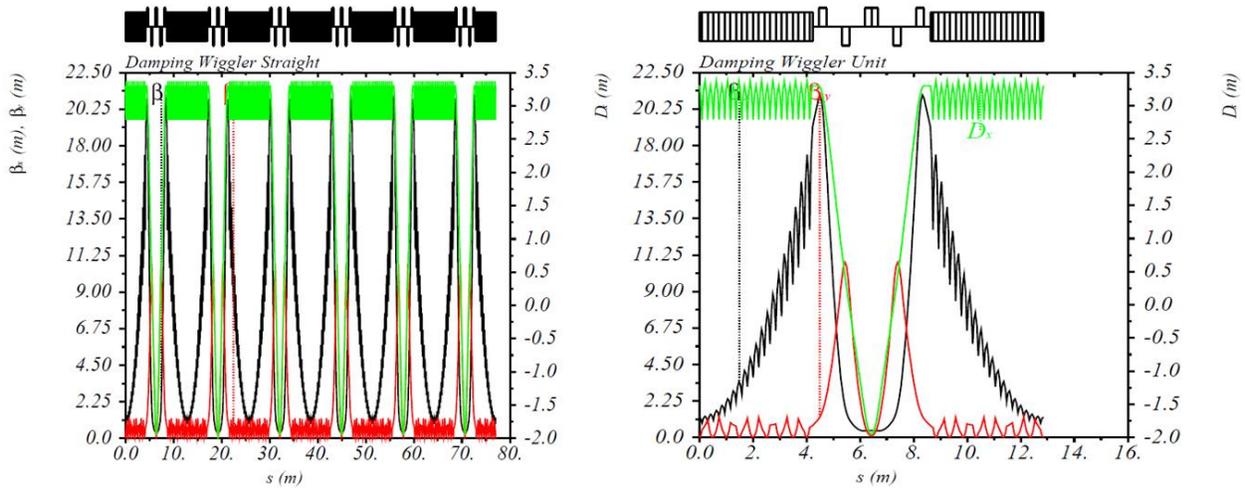

Figure 5. Optics functions of the whole damping wiggler section (left) and one damping wiggler unit (right) in the damping ring.

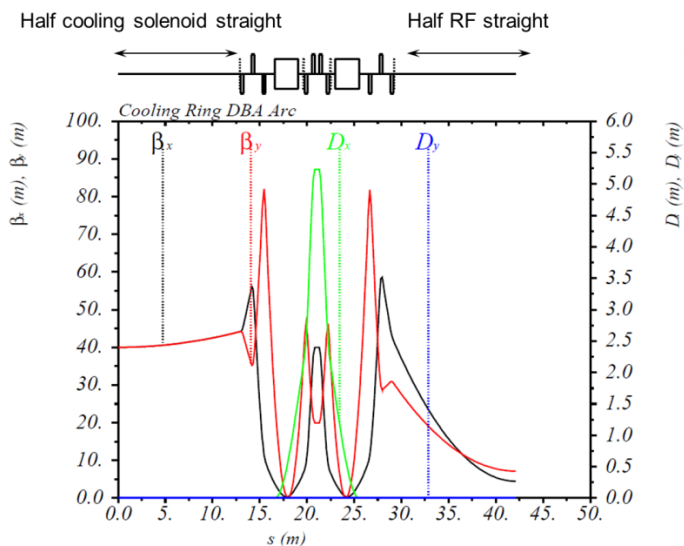
Figure 6: Optics functions of the cooling ring arcs.

## Discussion

Since an energy recovering SRF structure is needed in the storage-ring electron cooler, the excitation of High Order Modes (HOM) will be very strong with the required high beam current up to 1A. The HOMs will also be more disruptive to the beam due to the low beam energy and long damping time, especially for HOM induced instabilities such as multi-pass beam-breakup instabilities (BBU) and coupled bunch instabilities (CBI). We will analyze the HOM impedance requirements for these instabilities, and compare them with what has been achieved with existing cavity designs.

Any vertical emittance growth in the flat beam section will increase the Larmor radius inside the cooling solenoid and reduce the cooling efficiency. An alignment tolerance study is needed to find an upper limit of the vertical emittance growth.

## Conclusion

This paper describes the concept of a storage ring based cooler which may provide a high-current high-quality electron beam for efficient cooling of ion beams, particularly at ion energies above a few GeV. Such a cooler consists of two sections with significantly different energies: the cooling and damping sections. Energies at two sections are chosen for optimum cooling of the stored ion beam and optimum damping of the electron beam. A prototype linear optics design is presented and some potential issues are discussed. A consistent set of parameters of both cooler design and electron beam will be provided in the future.